\documentclass[fleqn,usenatbib]{mnras}

\usepackage{mathptmx}
\usepackage{subfigure}
\usepackage[T1]{fontenc}

\DeclareRobustCommand{\VAN}[3]{#2}
\let\VANthebibliography\thebibliography
\def\thebibliography{\DeclareRobustCommand{\VAN}[3]{##3}\VANthebibliography}

\usepackage{graphicx}	
\usepackage{amsmath}	
\usepackage{amssymb}	
\usepackage{color}

\newcommand{\mpc}{\, h^{-1} {\text{Mpc}}}
\newcommand{\kms}{\, \text{km}~\text{s}^{-1}}
\newcommand{\sag}{\textsc{\large{sag}}}
\newcommand{\roger}{\textsc{\large{roger}}}
\defcitealias{delosrios:2021}{dlR21}
\defcitealias{Coenda:2022}{C22}

\title[ROGER III]{Reconstructing Orbits of Galaxies in Extreme 
Regions (ROGER) III: galaxy evolution patterns in projected phase space around massive X-ray clusters}

\author[]{\parbox[t]{\textwidth}{
H\'ector J. Mart\'inez,$^{1,2}$\thanks{hjmartinez@unc.edu.ar}
Valeria Coenda,$^{1,2}$ 
Hern\'an Muriel,$^{1,2}$
Mart\'in de los Rios$,^{3,4}$
and Andr\'es N. Ruiz$^{1,2}$ 
}
\\
\\
$^{1}$Instituto de Astronom\'ia Te\'orica y Experimental (CCT C\'ordoba, CONICET, UNC), 
Laprida 854, X5000BGR, C\'ordoba, Argentina\\
$^{2}$Observatorio Astron\'omico, Universidad Nacional de C\'ordoba,
Laprida 854, X5000BGR, C\'ordoba, Argentina\\
$^{3}$ Departamento de F\'isica Te\'orica, Universidad Aut\'onoma de Madrid, 28049 Madrid, Spain\\
$^{4}$ Instituto de F\'isica Te\'orica, IFT-UAM/CSIC, C/ Nicolás Cabrera 13-15, Universidad Autónoma de Madrid, Cantoblanco, Madrid 28049, Spain
}

\date{Accepted XXX. Received YYY; in original form ZZZ}

\pubyear{2022}

\begin{document}
\label{firstpage}
\pagerange{\pageref{firstpage}--\pageref{lastpage}}
\maketitle

\begin{abstract}

We use the \roger\ code by de los Rios et al. to classify galaxies around a sample
of X-ray clusters into five classes according to their positions in the projected phase space 
diagram: cluster galaxies, backsplash galaxies, recent infallers, infalling galaxies,
and interlopers. To understand the effects of the cluster environment to the evolution of
galaxies, we compare across the five classes: stellar mass, specific star formation rate, size, and morphology.
Following the guidelines of Coenda et al., a separate analysis is carried out for red 
and blue galaxies. 
For red galaxies, cluster galaxies differ from the other classes, having a suppressed 
specific star formation rate, smaller sizes, and are more likely to be classified as 
ellipticals. Differences are smaller between the other classes, however 
backsplash galaxies have significantly lower specific star formation rates
than early or recent infalling galaxies.
For blue galaxies, we find evidence that recent infallers are smaller 
than infalling galaxies and interlopers, while the latter two are comparable in size.
Our results provide evidence that, after a single passage,
the cluster environment can diminish a galaxy's star formation, modify its morphology, 
and can also reduce in size blue galaxies.
We find evidence that quenching occurs faster than morphological transformation from spirals to ellipticals
for all classes. While quenching is evidently enhanced as soon as galaxies 
get into clusters, significant morphological transformations require galaxies to experience the
action of the physical mechanisms of the cluster for longer timescales.
\end{abstract}




\section{Introduction}
\label{sect:intro}

Galaxy clusters constitute the highest density virialised environment. They are characterised 
by a deep gravitational potential, hot and ionised intracluster gas, and thousands of galaxies.
Several physical properties of galaxies are known to vary systematically with environment: galaxy 
morphology (e.g. \citealt*{dressler80,Whitmore:1993,Dominguez:2001}; \citealt{Weinmann:2006,Bamford:2009,
Paulino-Afonso:2019}), 
colour (e.g. \citealt{Blanton05,Weinmann:2006,martinez06}; \citealt*{martinez08}), luminosity
(e.g. \citealt*{Adami:1998}; \citealt{Coenda:2006}), and the fraction of
star-forming galaxies 
(e.g. \citealt{Hashimoto,Mateus:2004,BlantonMoustakas:2009,Welikala:2008,Schaefer:2017,Coenda:2019}). 
In clusters, galaxies tend to become redder, more elliptical and with an older stellar 
population. In addition, they have less gas, less ongoing star formation and their spectra 
have less emission lines.

There are several mechanisms responsible for affecting galaxies inside clusters. 
Some of them induce gas depletion and consequently, cause the quenching of the star formation. 
One such mechanism is the ram pressure stripping (e.g. \citealt{GG:1972}; 
\citealt*{Abadi:1999,Book:2010, Vijayaraghavan:2015, Steinhauser:2016}), which removes the cold 
and warm gas of galaxies that are moving at high speeds through the hot ionised gas of the 
intracluster medium. This can cut off further gas cooling from the galaxy’s halo gas that fuels future 
star formation. Another mechanism that can remove the gas supply is the tidal stripping due to
the cluster potential (\citealt*{Zwicky:1951,Gnedin:2003a,Villalobos:2014}).
On the other hand, in the intermediate-density environments such as cluster outskirts and groups, 
mechanisms like galaxy-galaxy interaction, known as harassment, are more effective (e.g. 
\citealt{Moore:1996}; \citealt{Moore:1998}, \citealt{Gnedin:2003b}). This mechanism
can cause both: gas depletion, and morphological transformations. 
This tidal stripping from galaxy-galaxy encounters 
can produce the truncation of galaxies, in particular 
of disks, resulting in a spheroid-dominated galaxies (e.g. \citealt{Smith:2015}). 
Morphological evolution is 
thought to be governed principally by mergers (e.g. \citealt{Toomre:1977, Barnes:1992, DiMatteo:2007, 
Martin:2018}), frequent in groups, and much rarer in clusters given the high velocities of galaxies. 
Major mergers favour the formation of spheroidal systems \citep{Navarro:1994}, 
while gas-rich minor mergers could form massive disks \citep{Jackson:2022}.

During their lifetime, galaxies may live in different environments and experiment the effect of one or more of the mentioned mechanisms.
For instance, galaxies that are falling towards the clusters as part of a group (e.g. \citealt{McGee:2009,deLucia:2012,Wetzel:2013}; \citealt*{Hou:2014}) experiment different physical mechanisms than those that are falling from the field or through filament streams (e.g. \citealt{Colberg:1999}; \citealt*{Ebeling:2004,Martinez16}; \citealt{Rost:2020, Kuchner:2022}). In these falling processes, galaxies can also experiment several physical processes that lead to their transformation prior to entering the cluster, which is known as pre-processing (e.g. 
\citealt{Mihos:2004,Fujita:2004}). 

In the outskirts of clusters, 
in addition to galaxies that are being accreted, there are also backsplash galaxies 
\citep{Balogh:2000}. 
These galaxies have passed through the central regions of the clusters, experimenting the effects of denser environments, but now they are outside $R_{200}$,
the radius within which the mean density of a cluster is equal to 200 times the critical density of the Universe 
(e.g. \citealt{Mamon:2004}; \citealt*{Gill:2005,Rines:2006,Aguerri:2010,Muriel:2014, Kuchner:2022}).
Backsplash galaxies have undergone the effects of the cluster environment on their dive into
and out of the cluster, thus it is likely they have been affected in their properties,
but not to the point cluster galaxies have. This population should be characterised by 
physical properties intermediate between those of their field and cluster counterparts.

Several authors have used different definitions to classify galaxies that inhabit the clusters and its periphery, based on their position in the Projected Phase-Space Diagram (PPSD, e.g \citealt*{Mahajan:2011}; \citealt{Muzzin:2014, Muriel:2014, Jaffe:2015, OmanHudon:2016, Yoon:2017, Rhee:2017, Jaffe:2018, Pasquali:2019, Smith:2019}). This two-dimensional space combines the line-of-sight velocity relative to the cluster with the projected cluster-centric distance. Recently, 
\citet[hereafter dlR21]{delosrios:2021}  
present the \roger~code, which uses three different machine learning techniques to classify galaxies in, and around clusters, according to their projected phase-space position. They used a sample of massive and isolated galaxy clusters in the MultiDark Planck 3 \citep{klypin_mdpl2_2016}.
This code relates the two-dimensional PPSD position (2D) of galaxies to their orbital classification (3D). \citet[hereafter C22]{Coenda:2022} studied the properties of galaxies, as computed by the semi-analytic model of galaxy formation \sag~of \citet{cora_sag_2018}, 
using \roger\ to classify them into cluster galaxies, backsplash galaxies, galaxies that have recently fallen into a cluster, infalling galaxies, and interlopers. 
They find it is necessary to separate the galaxy populations in red and blue to perform a 
more reliable analysis of the results obtained out of 2D data. 
It is interesting to note that their predictions based on semi-analytical models applied to 
cosmological simulations are comparable with those found by \citet{Muriel:2014} 
in observations, namely backsplash galaxies are redder, form less stars and are older 
than infalling galaxies. On the other hand, \citet{Mahajan:2011} found that backsplash 
galaxies, whose absolute velocities are low, are more likely to be star forming than cluster (i.e. virialised) galaxies.

This paper is a natural continuation of both, \citetalias{delosrios:2021}, and 
\citetalias{Coenda:2022}, but using observational data instead of simulations. It is 
aimed at studying the properties of 
galaxies in and around bright X-ray galaxy clusters to deepen our understanding of the
effects of cluster environment upon galaxies. We rely on the \roger~code to classify galaxies
in the PPSD and the theoretical results found in C22 to interpret our results.
The present paper is organised as follows: we describe the samples of clusters and galaxies,
and the classification scheme based on \roger\ in Sec. \ref{sect:data};
the comparison of galaxy properties in and around clusters is carried out in Sec. 
\ref{sect:comp}; finally, we discuss our results and draw conclusions in Sec. 
\ref{sect:conclu}.

\begin{figure*}
\centering
  \subfigure{\includegraphics[width=\columnwidth]{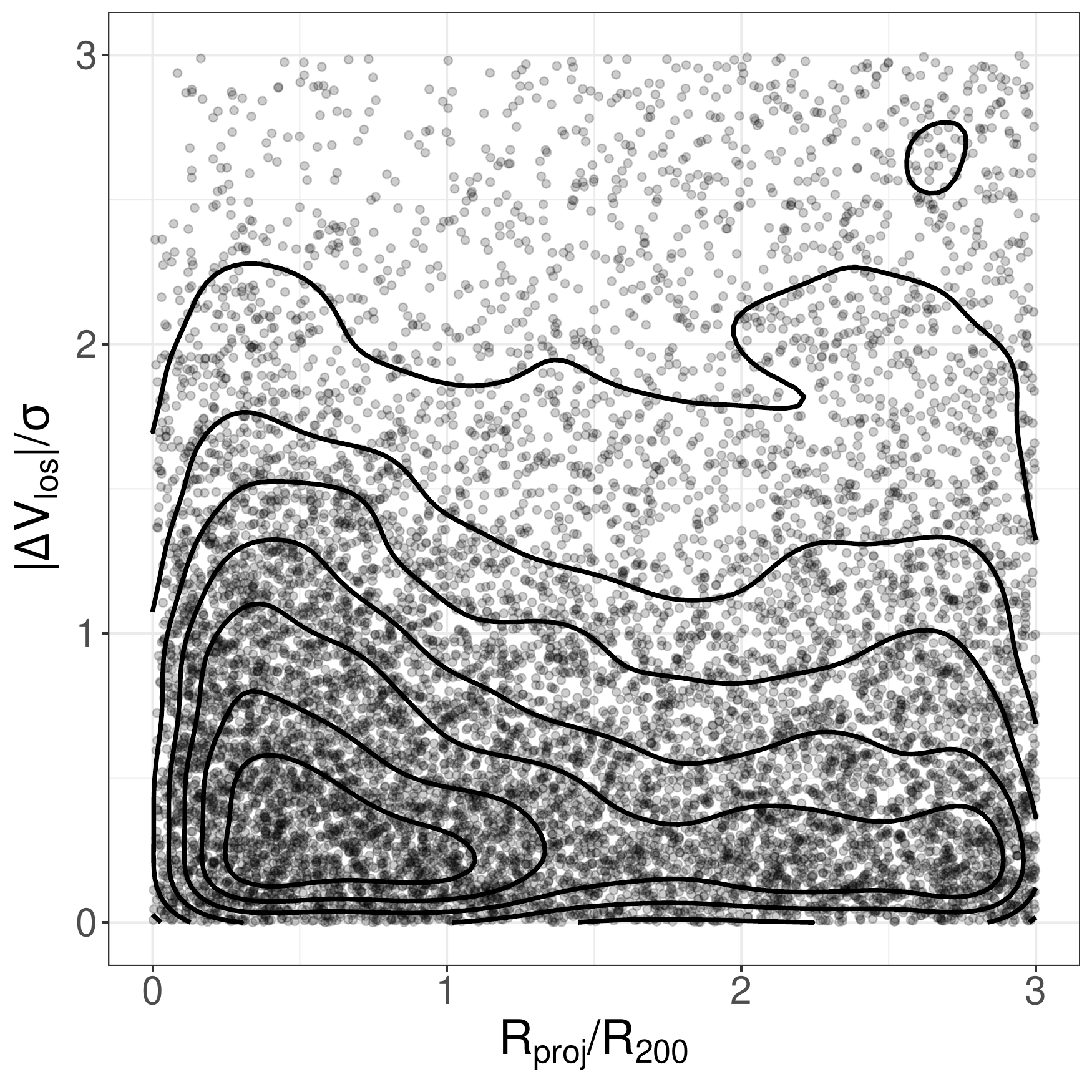}}  
  \subfigure{\includegraphics[width=\columnwidth]{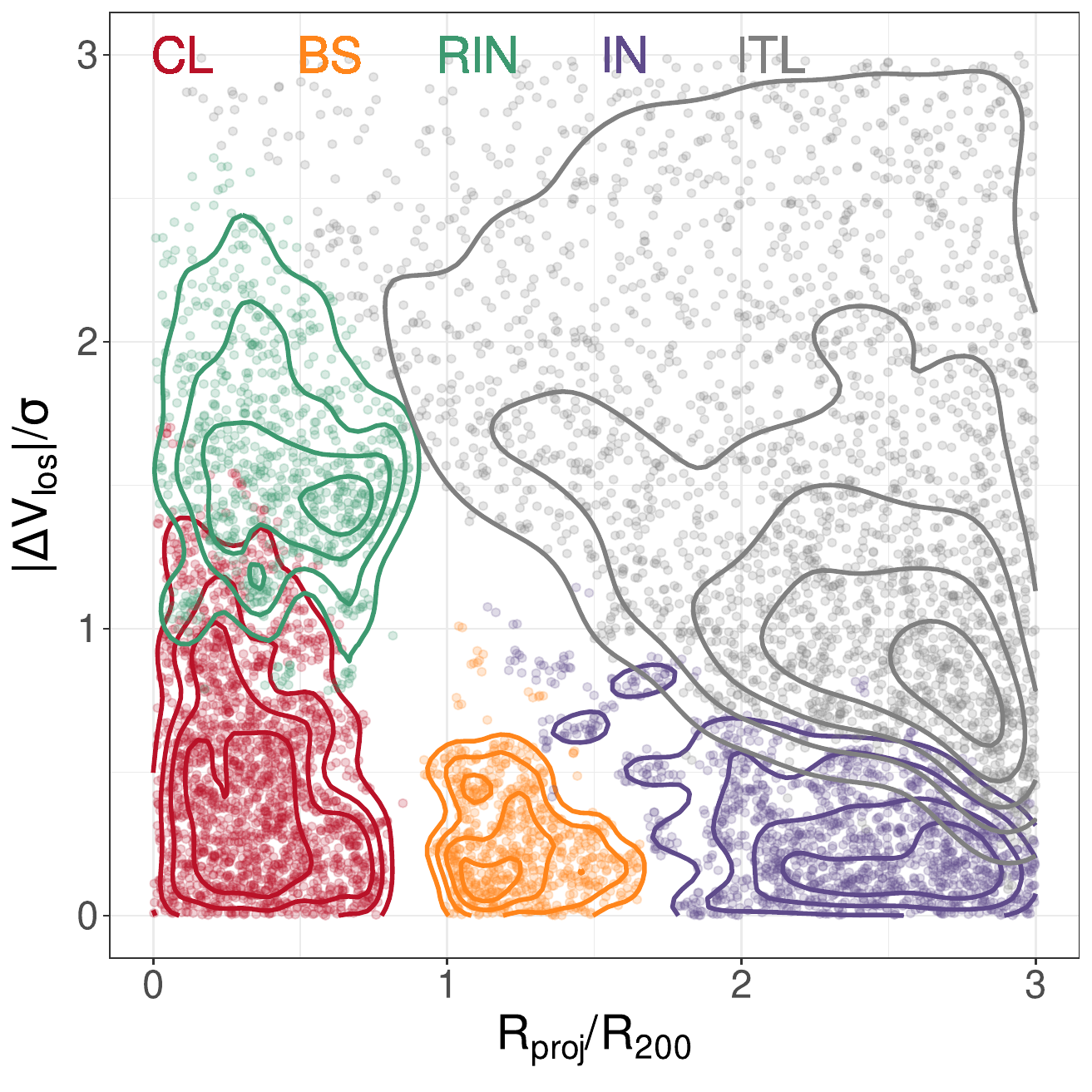}}
  \caption{Projected phase space diagram of SDSS galaxies around our sample
  of X-ray clusters. \textit{Left panel} shows the PPSD position of all galaxies
  in our initial sample. \textit{Right panel} shows the PPSD position of the galaxies in the
  left panel that were effectively classified into the predicted classes i-v (see text) according
  to \roger's probabilities and our threshold criteria. Colours denote the classification: red corresponds to CL, orange to BS, green to RIN, violet to IN, and gray to ITL.}
  \label{fig:ppsd}
\end{figure*}

\section{The samples} \label{sect:data}

\subsection{The sample of X-ray clusters}

Our sample of X-ray clusters has been drawn from two sources:
the C-P04-I sample of \citet{Coenda:2009}, and the C-B00-I sample of 
\citet{Muriel:2014}. The former was constructed from the ROSAT-SDSS Galaxy 
Cluster Survey of \citet{Popesso:2004}, and the latter from the
the Northern ROSAT All-Sky Galaxy Cluster Survey of \citet{Noras:2000},
and comprise 49 and 55 clusters, respectively, in the redshift range
$0.05\le z\le 0.14$. 
The authors
identified cluster members in two steps. First, they used a friends-of-friends 
(\citealt{H&G:1982}, hereafter \textit{fof}) algorithm that uses the linking 
parameters and modifications introduced by \citet{Diaz:2005}. Secondly, they 
performed eyeball examinations of the structures detected by the \textit{fof}
algorithm. From the redshift distribution of galaxies, they determined the galaxy
members from line-of-sight extension of each cluster.
By visually inspecting every cluster, the authors excluded systems that 
have two or more close substructures of similar size in the plane of the 
sky and/or in the redshift distribution. 

Using the galaxy members identified this way, the authors computed a number of cluster 
physical properties such as the line-of-sight velocity dispersion, virial radius, $R_{200}$, and virial mass. 
Clusters in the C-P04-I sample have median virial mass of 
$7\times10^{14}~M_{\odot}$, median $R_{200}$ of 
$1.77\mpc$, and median line-of-sight velocity dispersion of $715\kms$.
Clusters in the C-B00-I sample have median virial mass of 
$9\times10^{14}~M_{\odot}$, median $R_{200}$ of $2.03\mpc$, and median line-of-sight velocity dispersion of $820\kms$.


\subsection{The sample of galaxies}
All samples of galaxies used throughout this paper are drawn from the 
Main Galaxy Sample (MGS, \citealt{Strauss:2002}) of the Sloan Digital Sky 
Survey's (SDSS, \citealt{York:2000}) Seventh Data Release (DR7, \citealt{dr7}).
Throughout the paper, a flat cosmological
model is assumed with parameters $\Omega_0 = 0.3$, $\Omega_\Lambda=0.7$, and a Hubble’s constant $H_0 = 100\, h\, \kms \text{Mpc}^{-1}$. 
All magnitudes were corrected for Galactic extinction using the maps by
\citet*{sch98} and are in the AB system. Absolute magnitudes and galaxy colours were K-corrected
to $z=0.1$ using the method of \citet{Blanton:2003}.
The values of stellar mass and specific star formation rate for the galaxies in our sample have 
been extracted from the MPA-JHU DR7 release of spectra
measurements.  This catalogue provides, among other parameters,
stellar masses based on fits to the photometry following \citet{Kauff:2003} and \citet{Salim07} and star formation rates based
on \citet{Brinchmann:2004}. 
We consider the morphological classifications taken from the Galaxy Zoo Project 
\citep{Lintott:2008}, which provides likelihoods for every galaxy of being of a particular 
morphological type. In this work, we use the probabilities $P_{\rm S}$ (spiral) and $P_{\rm E}$ (elliptical) 
corrected after the de-biasing procedure of \citet{Lintott:2011}. 
We select all DR7 galaxies down to $r=17.77$ that are found around each cluster in our sample
within a redshift-space volume defined by projected distances $R_\text{proj}\leq 3\times R_{200}$, and
line-of-sight velocities $|\Delta V_\text{los}|\leq 3\times \sigma$, where $\sigma$ is the
line-of-sight velocity dispersion of the cluster.

\subsection{ROGER classification}\label{sect:roger}

When provided with a galaxy's PPSD position, namely, its projected distance to the 
nearest galaxy cluster in units of $R_{200}$, and its line-of-sight velocity relative to the
cluster in units of $\sigma$, 
\roger~computes the galaxy's probabilities of being any of the following five 
\textit{orbital classes}, as defined by \citetalias{delosrios:2021}:  
\begin{enumerate}
    \item Cluster galaxies (CL): galaxies that are satellites of the cluster. They have been 
    satellites for longer than 2 Gyr. The vast majority of them are inside $R_{200}$, with
    a few exceptions that are momentarily outside $R_{200}$ in their orbital motion. 
    \item Backsplash galaxies (BS): galaxies that have crossed $R_{200}$ exactly twice, the first
    in their way into the cluster, and the last in their way out. They are found outside
    $R_{200}$. These galaxies have the interesting feature of having experienced one passage
    through the cluster. They will most probably end up as CL in the future.
    \item Recent infallers (RIN): these are galaxies that are found within $R_{200}$, they have crossed
    it only once in their way in and no more than 2Gyr ago. These galaxies are
    experiencing the environmental action of the cluster environment for the first time in their
    lives. Some of them may become BS in the future.
    \item Infallers (IN): these galaxies have been outside $R_{200}$ their whole lifetimes, they are falling into the cluster, i.e., they have negative radial velocities relative
    to the cluster.
    \item Interlopers (ITL): these galaxies have also been outside $R_{200}$ for their whole lifetimes,
    but they are not approaching the cluster, and they are not related to the cluster in any physical way. They appear in the PPSD only due to projection effects.
\end{enumerate}

\begin{figure}
\includegraphics[width=1\columnwidth]{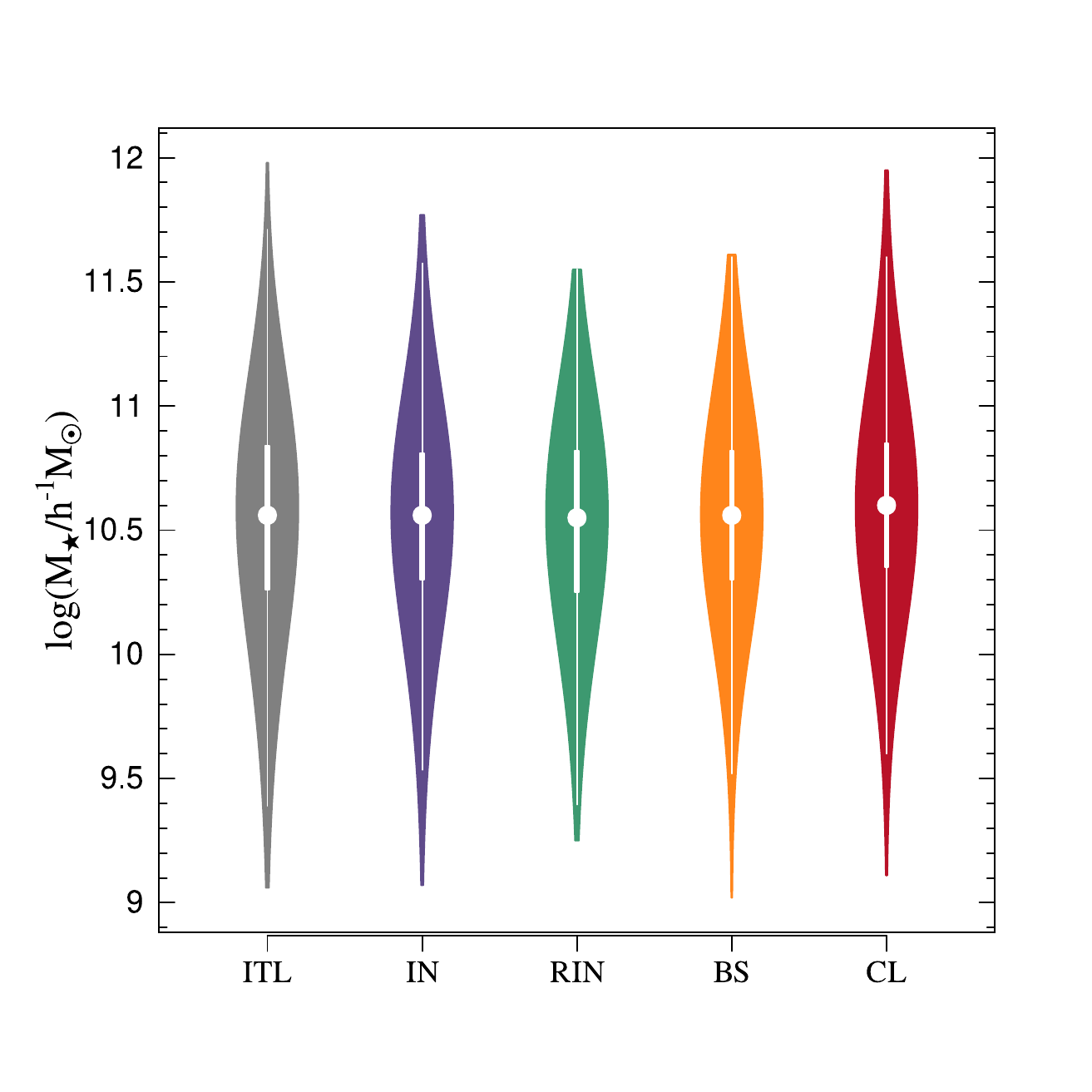}
\caption{Violin plot of the stellar mass distributions of galaxies in the 5 predicted classes. colours 
are as in Fig. \ref{fig:ppsd}. White circles are median values, 
while the thick white bars are the 25 and 75\% percentiles.}
\label{fig:mass}
\end{figure}

\begin{figure*}
\includegraphics[width=2\columnwidth]{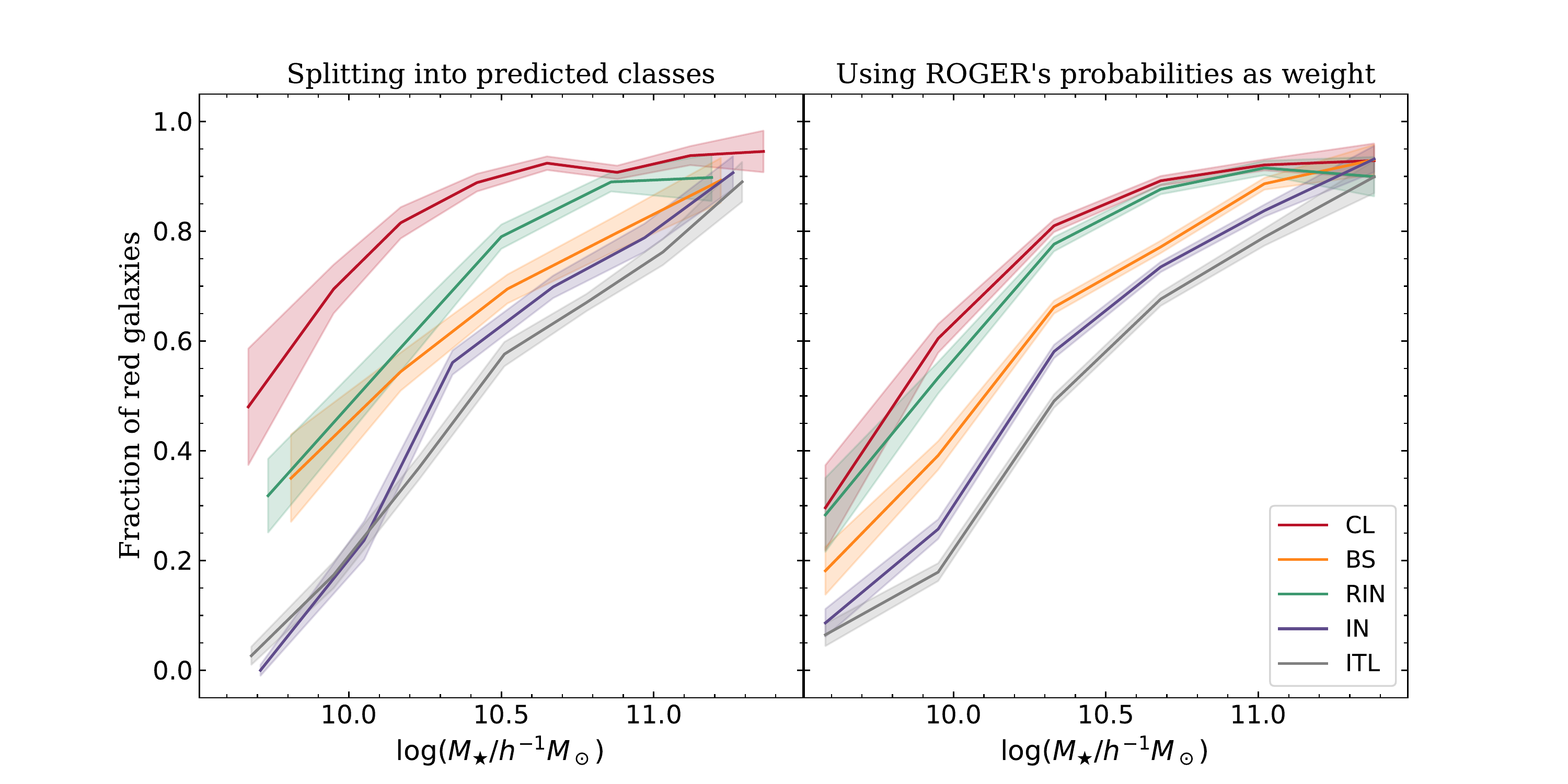}
\caption{\textit{Left panel:} fraction of red galaxies as a function of stellar mass
for our predicted classes. \textit{Right panel:} fraction of red galaxies using \roger's probabilities
as weight, in this case galaxies are not split into the 5 predicted classes but they all contribute
to all curves. Errors were computed using the bootstrap resampling
technique.}
\label{fig:frac}
\end{figure*}
We show in the left panel of Fig. \ref{fig:ppsd} the PPSD position of all galaxies in our sample.
\roger~ was trained by \citetalias{delosrios:2021} to compute the probabilities for a galaxy to be
of classes i-v using three Machine Learning techniques: K-Nearest Neighbours (KNN), Support 
Vector Machines, and Random Forest. In the light of the \citetalias{delosrios:2021}'s results, we
adopt the trained KNN technique to compute class probabilities for our galaxies.
Then, we use the \citetalias{Coenda:2022} criteria to classify our galaxies: a galaxy will be considered 
as belonging to a particular class if, among the five probabilities computed the highest one 
corresponds to that class, and also, if that probability is higher than a certain threshold.
These thresholds have been quoted in Table \ref{tab:sample} and were chosen by 
\citetalias{Coenda:2022} as 
a compromise between precision and sensitivity. 
Hereafter we will refer to these classes as \textit{predicted classes}.

The adoption of probability thresholds determines that not all galaxies in our sample 
meet both criteria as to end up classified into one of the five predicted classes. 
These galaxies are from now on excluded from our analysis. 
In the right panel of Fig. \ref{fig:ppsd} we show with different colours the position in the
PPSD of those galaxies that meet our classification criteria. 
In contrast with the left panel of the figure, in the right panel there are zones empty 
of galaxies. In these zones, galaxies of the five orbital classes defined above strongly
overlap, thus it is more unlikely for \roger~to assign a galaxy located there a high
probability to be of a particular class. Adopting thresholds gives more reliable 
results at the price of loosing objects. The resulting number of galaxies
in each predicted class is quoted in Table \ref{tab:sample}.

Classifying galaxies into predicted classes is not the only way of using \roger's
probabilities, another possibility is using the probabilities as weights. The user can
compute all desired statistics weighting all galaxies in their sample with the
chosen probability.

\begin{table}
    \caption{Thresholds used in the PPSD classification and the resulting number of galaxies 
    in each predicted class}
    \label{tab:sample}
    \begin{tabular}{l|c|c|c|c|c}
        \hline
        Class & ITL & IN & RIN & BS & CL \\
        \hline
        Threshold$^a$           & $0.15$  & $0.54$ & $0.37$ & $0.48$ & $0.40$ \\
        Number of galaxies      & $2541$ & $1433$  & $773$  & $711$ & $1907$ \\
        \hline
        $^a$ \citet[Table 1]{Coenda:2022}
    \end{tabular}
\end{table}

\begin{figure*}
\centering
\includegraphics[width=17 cm]{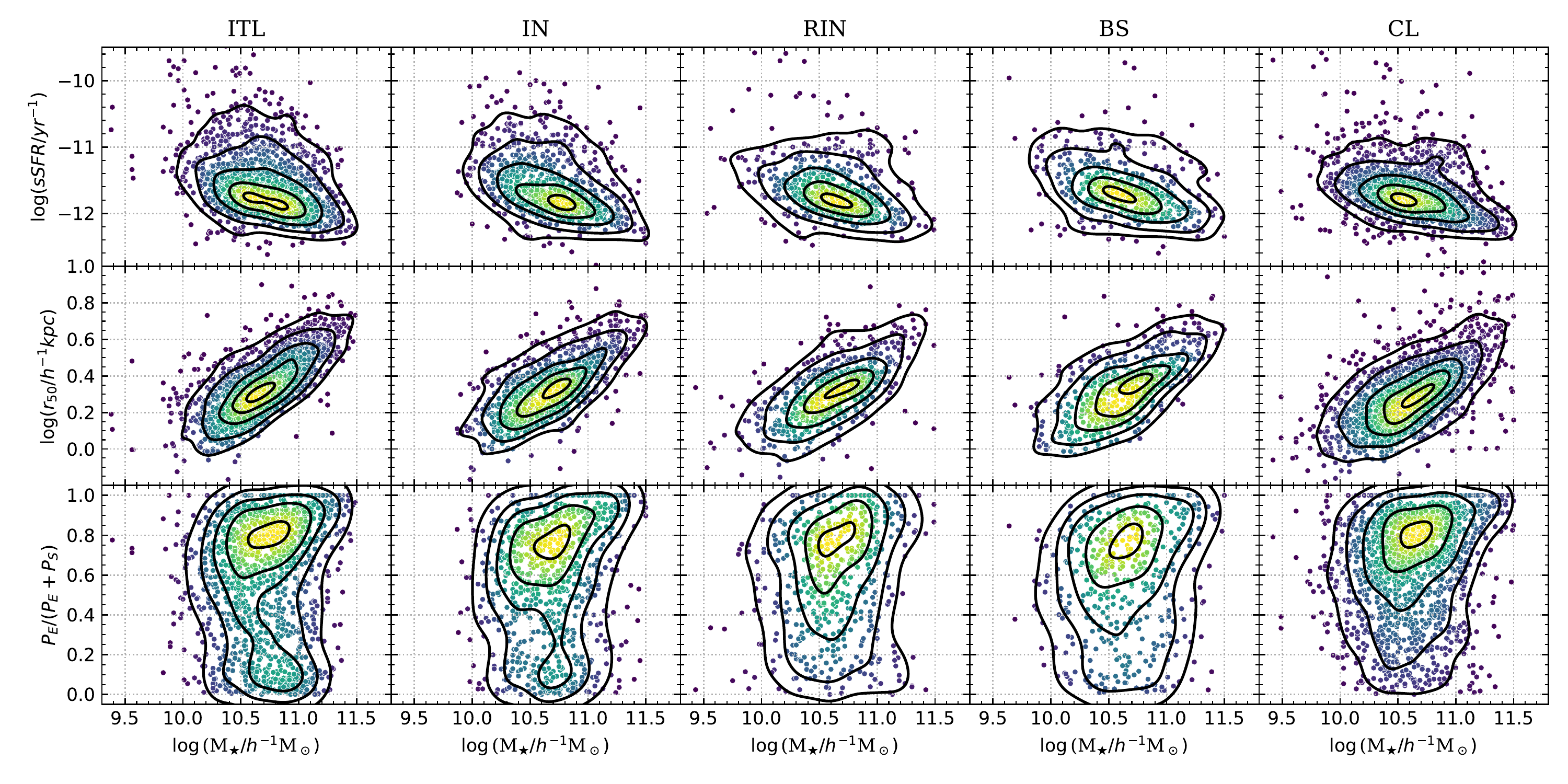}
\caption{Upper panels: specific star formation rate as a function of stellar mass for red sequence 
galaxies classified into the five predicted classes, as quoted above each panel. Dot colour 
denotes local numerical density in the $\log(M_{\star})-\log(\mathrm{sSFR})$ plane. 
Contours enclose 90, 75, 50, 25 and $5\%$ of the datapoints 
from the outermost to the innermost.
Middle panels: similar to the upper panels but showing $\log(r_{50})$ 
as a function of stellar mass.
Lower panels: similar to the upper and middle panels but showing 
$P_{\rm E}/(P_{\rm E}+P_{\rm S})$ as a function of stellar mass.
}
\label{fig:scatterR}
\end{figure*}

\begin{figure*}
\centering
\includegraphics[width=17 cm]{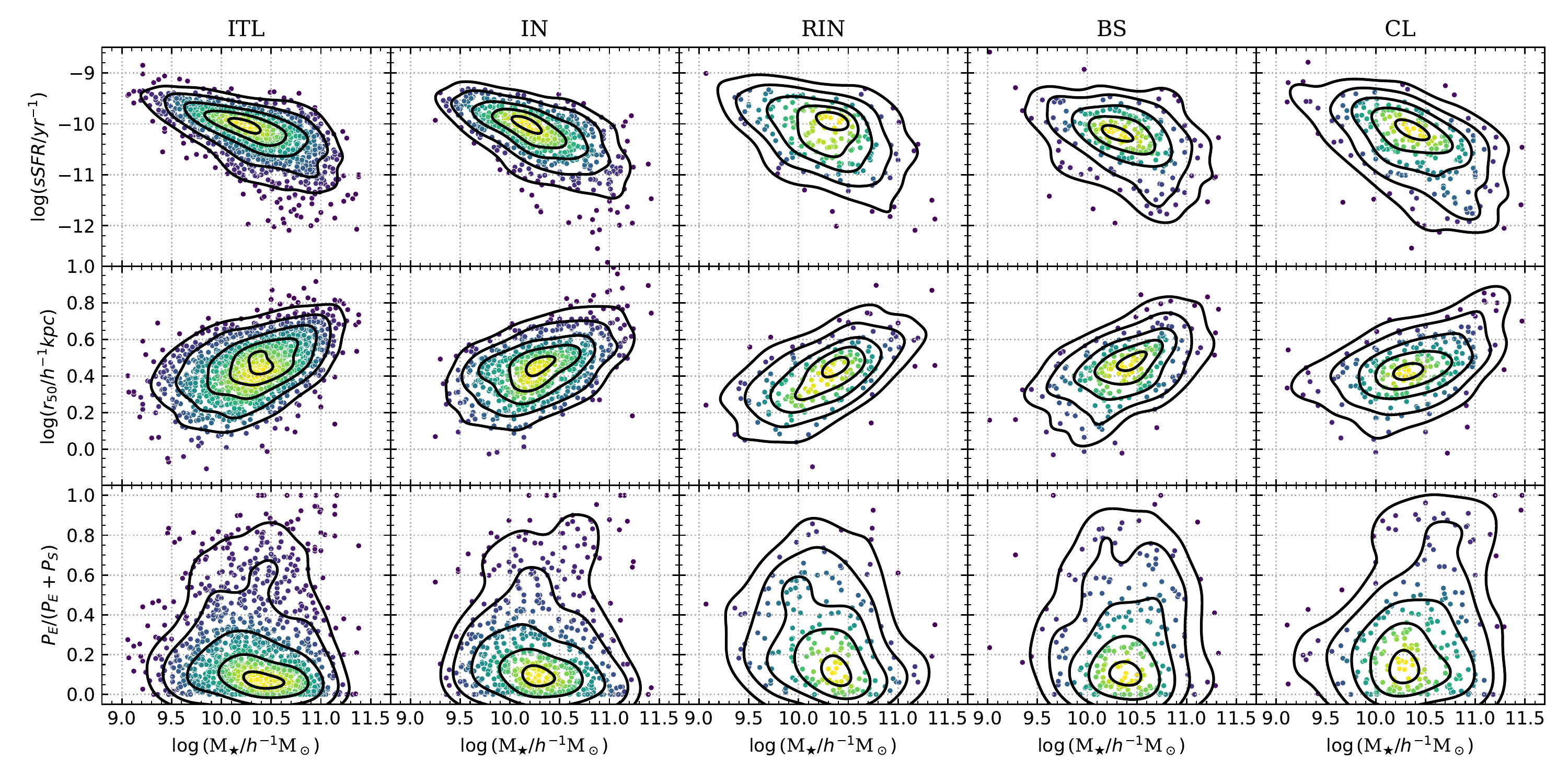}
\caption{
Similar as Fig. \ref{fig:scatterR} but for blue cloud galaxies.
}
\label{fig:scatterB}
\end{figure*}

\begin{figure*}
\centering
\includegraphics[width=19 cm]{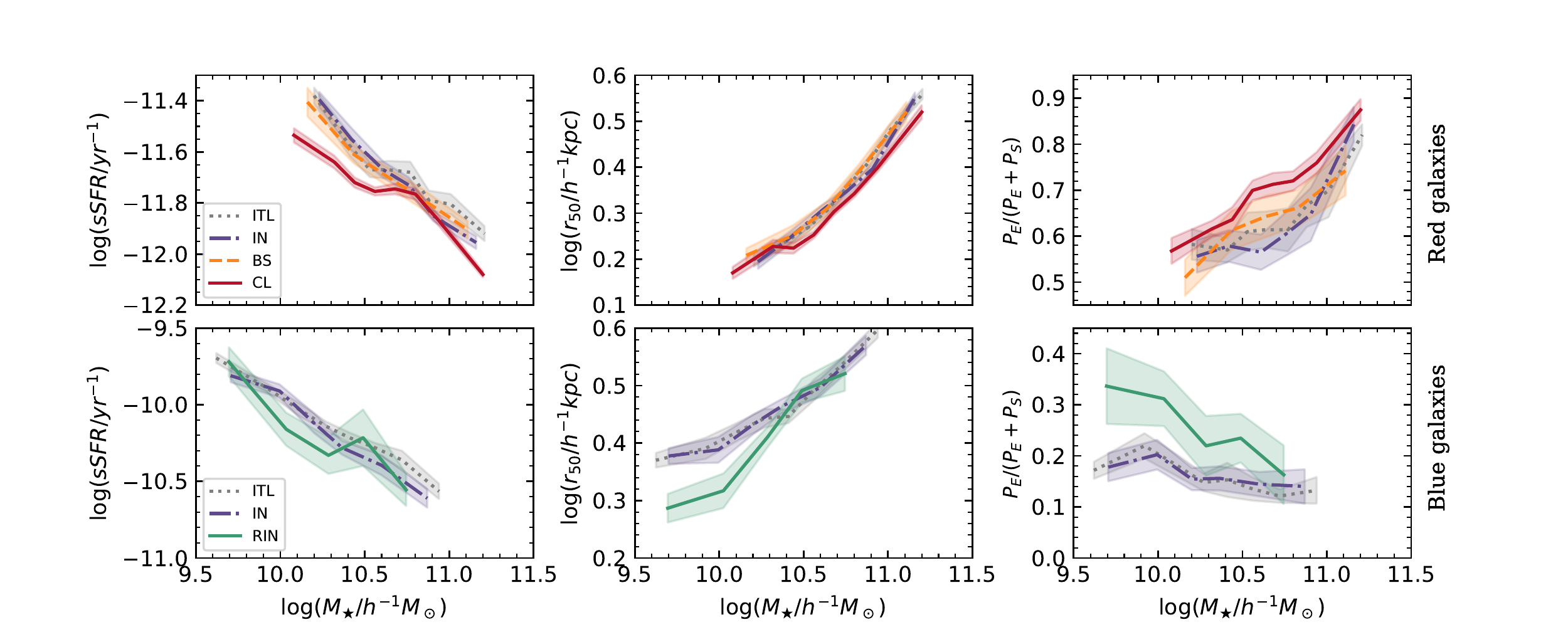}
\caption{The medians of $\log(\mathrm{sSFR})$ (left panels), $\log(r_{50})$ (centre panels), 
and $P_{E}/(P_{E}+P_{S})$ as a function of stellar mass. 
Upper panels consider red sequence galaxies, while lower panels blue cloud
galaxies. 
For each class and sequence, bins have been chosen to have the same number of galaxies.
Red sequence: ITL: 182; IN: 147; BS: 99; CL: 223. Blue sequence: ITL: 157; IN: 100; 
RIN: 38.
$x-$axis positions mark the median stellar mass within each bin. 
Error bars were computed using the bootstrap resampling technique.}
\label{fig:medians}
\end{figure*}

\section{A comparison of physical properties of galaxies in the predicted classes}
\label{sect:comp}

In this section, we compare a number of properties of galaxies in the different predicted
classes: stellar mass ($M_{\star}$), specific star formation rate (sSFR), 
the radius enclosing half the $r-$band Petrosian flux ($r_{50}$), 
and morphology by means of the Galaxy Zoo's 
probability of being elliptical ($P_{\rm E}$) or spiral ($P_{\rm S}$).  
Since most galaxy properties correlate with stellar mass (e.g. \citealt{Kauff:2003,Conselice:2006}), a fair comparison across 
predicted classes requires that they have similar stellar mass distributions. This is the case of
our subsamples as can be seen in Fig. \ref{fig:mass}, where only subtle differences can be 
observed.

\citetalias{Coenda:2022} showed how careful one should be when drawing conclusions from comparisons of 
galaxy 
properties in the five predicted classes. As shown by \citetalias{delosrios:2021}, overlapping between 
orbital 
classes on the PPSD follow clear patterns, they preferentially occur between CL and RIN, 
and between BS and IN. Accordingly, when studying the properties of the predicted 
classes, it is of central importance to carry out separate analysis for blue and red galaxies. 
Let us consider first what should be handled with spacial care: the blue cloud of predicted CL 
galaxies is made up mostly of misclassified real RIN; conversely, the red sequence of predicted 
RIN is formed out of misclassified CL; the blue cloud of BS is significantly
enhanced in numbers by misclassified IN. On the other hand, \citetalias{Coenda:2022}
showed that the properties
of the following populations of galaxies are recovered reliably by our classification scheme: 
the red sequence of CL, BS and IN; the blue cloud of RIN and IN. Furthermore, the properties 
of both, the blue, and the red populations of ITL galaxies are recovered reliably.  
In this paper, to separate galaxies into blue and red, we use the 
$^{0.1}(u-r)$ colour cut defined by \citet[Sec. 3.2.2]{ZM11}, which is
a quadratic polynomial in absolute $^{0.1}r-$band magnitude given by $P(x) = -0.02x^2 - 0.15x + 2.46$, where $x=M_{r}^{0.1}-5\log(h)+20$.

In the left panel of Fig. \ref{fig:frac}, we show the fraction of red galaxies 
as a function of galaxy stellar mass for galaxies in our five predicted classes. 
Clear trends with mass and predicted class are seen. It is noticeable that RIN show a fraction 
of red galaxies larger than that of BS, which can be counterintuitive. Nonetheless,
as shown by \citetalias{delosrios:2021} and \citetalias{Coenda:2022}, galaxies in 
this predicted
class are highly contaminated by CL, which are predominantly red galaxies.
The comparison of the fraction of red galaxies in the predicted classes BS and IN resembles
the findings by \citet[their Fig. 8]{Muriel:2014}.
We show in the right panel of Fig. \ref{fig:frac} the fraction of red galaxies
in the five classes using \roger\ probabilities as weights. In this case all galaxies
in our original sample contribute to the five curves shown. This is an example of an  
alternative use of \roger\ probabilities. Qualitatively, the results in this panel are
consistent with those in the left panel. Statistical errors are smaller for some classes 
given the larger number of galaxies involved. However, CL and RIN cross-contamination
is more clearly evidenced. We prefer the use of thresholds to deal with purer samples
for a better comparison across classes.

In Fig. \ref{fig:scatterR} we show the scatter plots of sSFR (upper panels), $r_{50}$ 
(middle panels), and the fraction $F_{\rm E}\equiv P_{\rm E}/(P_{\rm E}+P_{\rm S})$ (bottom panels) as a 
function of stellar mass for 
red sequence galaxies in the five predicted classes.
We use this single morphological quantity $F_{\rm E}$ to focus our study into how likely to be elliptical/spiral 
each galaxy is, putting aside other morphological classes.
Similarly, in Fig. \ref{fig:scatterB} we show the corresponding scatter plots for blue cloud galaxies.
In both figures, we show the contours that enclose 95, 75, 50, 25, and 5\% of data points.
Fig. \ref{fig:medians} compares the medians of the scatter plots shown in Figs. \ref{fig:scatterR}
and \ref{fig:scatterB} as a function of
stellar mass for those sequences and predicted classes that can be fairly compared, i.e.,
the red sequence of ITL, IN, BS and CL, and the blue cloud of ITL, IN and RIN.

\begin{table*}
    \centering
    \caption{The mean difference per mass bin, $\Delta y_{\rm AB}$, and the rejection probability
    $\mathcal{R}_\text{p}$, of the null hypothesis of two predicted classes, $A$ and $B$, being 
    drawn from the same underlying population (see Appendix \ref{test_lares} for details). In the
    main body of the table, the computations are performed over the whole mass range.
    For a few cases in which differences in the trends shown in 
    Fig. \ref{fig:medians} are particularly strong, we have repeated the computations restricting
    the mass range to the low mass end
    ($\log(M_{\star}/h^{-1}M_{\odot})<10.3$) for blue galaxies. Values of $\mathcal{R}_{\rm
    p} > 0.99$ are highlighted in boldface and those between $0.95$ and $0.99$ are highlighted 
    in italics.}
    \label{tab:rejprob}
    \begin{tabular}{l|c|c|c|c|c|c|c|c|c|c|c|c}
        \hline 
         &
         \multicolumn{4}{c}{$y=\log(\mathrm{sSFR}/\mathrm{yr}^{-1})$} &  
         \multicolumn{4}{c}{$y=\log(r_{50}/h^{-1}\mathrm{kpc})$} &
         \multicolumn{4}{c}{$y=P_{\rm E}/(P_{\rm E}+P_{\rm S})$}\\
        \hline
        Classes & 
        \multicolumn{2}{c}{Red} & 
        \multicolumn{2}{c}{Blue} & 
        \multicolumn{2}{c}{Red} & 
        \multicolumn{2}{c}{Blue} & 
        \multicolumn{2}{c}{Red} & 
        \multicolumn{2}{c}{Blue} \\
         $A-B$ & 
        $\Delta y_{\rm AB}$ & $\mathcal{R}_\text{p}$ &
        $\Delta y_{\rm AB}$ & $\mathcal{R}_\text{p}$ &
        $\Delta y_{\rm AB}$ & $\mathcal{R}_\text{p}$ &
        $\Delta y_{\rm AB}$ & $\mathcal{R}_\text{p}$ &
        $\Delta y_{\rm AB}$ & $\mathcal{R}_\text{p}$ &
        $\Delta y_{\rm AB}$ & $\mathcal{R}_\text{p}$ \\
        \hline
        ITL--IN  & $0.041$  & {\it 0.977} & $0.043$  & {\it 0.973} &
                   $0.005$  & $0.788$     & $-0.001$ & $0.542$     &
                   $-0.010$ & $0.752$     & $-0.009$ & $0.765$     \\
        ITL--RIN & $0.103$  & {\bf 1}     & $0.055$  & $0.943$     &
                   $0.014$  & {\it 0.980} & $0.033$  & {\bf 0.998} &
                   $-0.056$ & {\bf 1}     & $-0.055$ & {\bf 0.997} \\
        ITL--BS  & $0.083$  & {\bf 1}     & $0.064$  & $0.853$     &
                   $0.004$  & $0.695$     & $0.019$  & {\it 0.958} &
                   $-0.025$ & $0.933$     & $-0.052$ & {\bf 0.999} \\
        ITL--CL  & $0.141$  & {\bf 1}     & $0.165$  & {\bf 1}     &
                   $0.031$  & {\bf 1}     & $0.024$  & {\it 0.989} &
                   $-0.093$ & {\bf 1}     & $-0.054$ & {\bf 0.998} \\
        IN--RIN  & $0.065$  & {\bf 0.997} & $0.031$  & $0.776$     &
                   $0.010$  & $0.918$     & $0.034$  & {\bf 0.997} &
                   $-0.046$ & {\bf 0.998} & $-0.047$ & {\it 0.985} \\
        IN--BS   & $0.051$  & {\it 0.981} & $0.028$  & $0.780$     &
                   $0.000$  & $0.525$     & $0.020$  & {\it 0.967} &
                   $-0.018$ & $0.849$     & $-0.041$ & {\it 0.975} \\
        IN--CL   & $0.107$  & {\bf 1}     & $0.115$  & {\bf 0.999} &
                   $0.027$  & {\bf 1}     & $0.024$  & {\it 0.986} &
                   $-0.085$ & {\bf 1}     & $-0.041$ & {\it 0.983} \\
        RIN--BS  & $-0.017$ & $0.757$     & $0.026$  & $0.693$     &
                   $-0.009$ & $0.864$     & $-0.010$ & $0.745$     &
                   $0.025$  & $0.917$     & $0.003$  & $0.541$     \\
        RIN--CL  & $0.041$  & {\it 0.982} & $0.102$  & {\it 0.955} & 
                   $0.018$  & {\bf 0.993} & $-0.007$ & $0.685$     & 
                   $-0.040$ & {\bf 0.998} & $0.000$  & $0.502$     \\
        BS--CL   & $0.056$  & {\bf 0.998} & $0.097$  & {\it 0.960} & 
                   $0.026$  & {\bf 1}     & $0.007$  & $0.691$     & 
                   $-0.065$ & {\bf 1}     & $-0.006$ & $0.589$     \\    
        \hline
        Low mass end& \\
        ITL--IN  & $-$      & $-$     & $0.027$  & $0.826$ & 
                   $-$      & $-$     & $0.003$  & $0.619$ &
                   $-$      & $-$     & $-0.004$ & $0.590$ \\
        ITL--RIN & $-$      & $-$     & $0.067$  & $0.921$ & 
                   $-$      & $-$     & $0.068$  & {\bf 1} &
                   $-$      & $-$     & $-0.097$ & {\bf 1} \\
        IN--RIN  & $-$      & $-$     & $0.039$  & $0.781$ & 
                   $-$      & $-$     & $0.065$  & {\bf 1} &
                   $-$      & $-$     & $-0.093$ & {\bf 1} \\
        \hline
    \end{tabular}
\end{table*}

To tell whether the underlying distributions actually differ, we complement our 
analysis by means of estimating the rejection probability of the null hypothesis in 
which the distribution of points in the plane $\log(M_{\star})-y$ (where $y$ can be 
$\log(\mathrm{sSFR}/\mathrm{yr}^{-1})$, $\log(r_{50}/h^{-1}\mathrm{kpc})$, or $F_{\rm E}$) for each 
pair of sets of galaxies we are comparing are drawn from the same underlying 
distribution following \citet{Muriel:2014}. 
They propose a shuffling method using the statistic of the mean over mass bins
of the mean differences per mass bin, $\Delta y_{\rm AB}$
(see Appendix \ref{test_lares} below for 
details). The quantities $\Delta y_{\rm AB}$ and the corresponding rejection probabilities 
$\mathcal{R}_\mathrm{p}$ for the properties analysed here are quoted in
Table \ref{tab:rejprob}.\\

For red sequence galaxies we find:
\begin{enumerate}
\item
For the sSFR, moving from ITL towards CL in the upper panels of Fig. \ref{fig:scatterR}, 
there is a progressive contraction of the outer contours that gradually enclose less galaxies 
with higher values of sSFR. 
This global trend includes RIN, which, may be non negligibly 
contaminated by CL as pointed out by \citetalias{Coenda:2022}. 
When comparing the medians in the upper left panel of Fig. \ref{fig:medians}, it is 
clear that CL distinguish from the other predicted classes. No noticeable differences are seen 
between the medians of BS, IN and ITL. 
Rejection probabilities associated with the comparison of $\log(\mathrm{sSFR})$ vs. $\log(M_{\star})$, 
for red sequence galaxies are quoted in the third column of Table \ref{tab:rejprob}.
Unsurprisingly, on the one hand, CL differs with BS, IN, and ITL at significance levels $>3\sigma$,
on the other hand, in the RIN-CL comparison the rejection reaches a lesser significant 
($>2\sigma$) level, which is expected from the results of \citetalias{Coenda:2022}.
Of more interest is the comparison between BS and the other classes, they differ significantly
only from ITL at $>3\sigma$ level. 
\item
Regarding $r_{50}$ as a function of stellar mass, the inspection of middle panels of 
Fig. \ref{fig:scatterR}, the upper central panel of Fig. \ref{fig:medians}, and the seventh 
column of Table \ref{tab:rejprob} leads to the conclusion that the only significant difference
is found between CL and the other four classes. 
CL galaxies are typically smaller than
the other classes at a fixed stellar mass. This trend has been reported by other authors  
(e.g. \citealt{Coenda:2009, Cebrian:2014, Matteuzzi:2022})
\item
As for the morphological parameter $F_{\rm E}$, lower panels of Fig. 
\ref{fig:scatterR} show a decreasing fraction of low $F_{\rm E}$ galaxies as we move from ITL 
to CL. 
Medians in the upper right panel of Fig. \ref{fig:medians} suggest the only predicted
class that clearly distinguishes from the other are CL galaxies.
This is confirmed by the rejection probabilities quoted in the eleventh column of
Table \ref{tab:rejprob}
\end{enumerate}

For blue cloud galaxies we find:
\begin{enumerate}
\item
For $\log(\mathrm{sSFR})$ vs. $\log(M_{\star})$, in the upper panels of Fig. \ref{fig:scatterB} 
the tight sequence of star forming galaxies seen for ITL broadens as we move 
rightwards towards CL galaxies to include an increasing fraction of lower star forming galaxies. 
Recall that a non negligible contamination is expected for blue CL and BS galaxies.
The medians shown in the lower left panel of Fig. \ref{fig:medians}, while not significantly
different, suggest decreasing values of sSFR with increasing stellar mass (e.g. \citealt{Belfiore:2018,Spindler:2018, coenda18}), when considering ITL, IN, and RIN, thus sorted. These results are in agreement with those obtained by C22 in \sag: RIN are more passive than IN, followed by ITL galaxies. 
As for the rejection test, CL galaxies stand out once more as clearly different from the other
classes, even though a significant degree of contamination is to be expected for them.
At a $2\sigma$ differentiation level we have ITL vs. IN. The lower left panel of Fig. \ref{fig:medians} suggests a greater difference between RIN and
the classes ITL and IN at the low mass end, $\log(M_{\star}/h^{-1}M_{\odot})< 10.3$. We
compute the rejection test for these classes at this mass range (see lower rows of Table 
\ref{tab:rejprob}) not finding different results.
\item
For $\log(r_{50})$ vs. $\log(M_{\star})$, in the middle panels of Fig. \ref{fig:scatterB},
a decrease of size from IN to RIN occurs for low mass galaxies. In the 
comparison of medians of Fig. \ref{fig:medians}, 
RIN appear to be smaller than IN and ITL for $\log(M_{\star}/h^{-1}M_{\odot})\lesssim 10.4$.
According to the computed $\mathcal{R}_{\text p}$ (ninth column of Table
\ref{tab:rejprob}), blue RIN are significantly ($>3\sigma$)
different (more compact) from IN and ITL, but no so with respect to CL and BS, both of which are expected 
to be contaminated, the former by RIN and the latter by IN.
The rejection test applied to the low mass in the comparison between RIN and the classes 
ITL and IN give even greater differences.
\item
For $F_{\rm E}$ vs. $\log(M_{\star})$, in the lower panels of Fig. \ref{fig:scatterB},
the higher concentration towards lower values of $F_{\rm E}$ seen for ITL progressively dilutes 
as we move towards CL.
As for the comparison of medians, the lower right panel of Fig. \ref{fig:medians} shows
a clear distinction between RIN and the predicted IN and ITL classes.
The computed $\mathcal{R}_{\text p}$ (thirteenth column of Table \ref{tab:rejprob})
indicate that the highest level of rejection takes place in the comparison of ITL vs RIN, 
at a $3\sigma$ level. Once again, restricting the rejection test to the low mass end give 
greater differences between RIN and the classes ITL and IN.
All these results suggest that blue galaxies that have recently fallen into a cluster
may suffer fast morphological transformations.
\end{enumerate}

\begin{figure*}
\includegraphics[width=2.1\columnwidth]{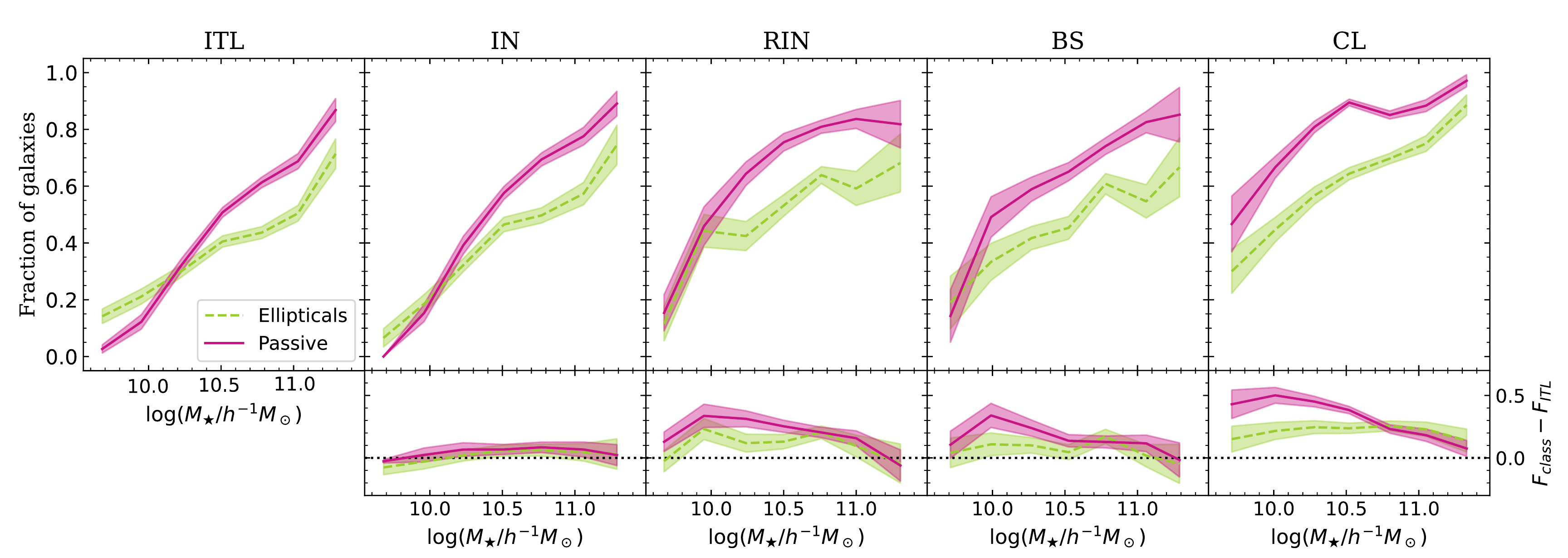}
\caption{The fraction of early type (light green dashed lines), and passive galaxies 
(purple continuous lines) as a function of stellar mass
for our predicted classes. Errors were computed using the bootstrap resampling
technique. In the lower panels of columns 2--5 we show the difference of the 
fraction of early and of passive galaxies relative to the corresponding fractions
for the ITL class.} 
\label{fig:comp_frac}
\end{figure*}

A few high rejection probability values quoted in Table \ref{tab:rejprob}, 
$\mathcal{R}_\mathrm{p} > 0.95$, 
provide some hints to help us understand the global picture of 
the evolution of galaxies in clusters. 
One
notices significant steps in galaxy evolution in clusters,
i.e., the order ITL-IN-RIN-BS-CL. For blue galaxies the steps from ITL to IN 
and BS to CL affect the sSFR; IN to RIN strongly affects the size and the morphological mix. 
For red
galaxies, again ITL to IN affects sSFR, but steps IN to RIN and BS to ICL reduce sSFR
much more significantly;
at high levels of significance, 
BS to CL affects the size and morphological mix, and IN to RIN the morphological mix. It is also
worth noticing that there are signs of star formation quenching in the currently infalling galaxies
relative to interlopers for both, blue and red galaxies.

These results suggest that diving into a cluster's inner regions for 
the first time (IN to RIN) results in further quenching of (already quenched) red galaxies, on 
size decrease or compaction \citep{Zolotov:2015} for blue galaxies and morphological mix for blue 
and especially red galaxies. 
The BS to CL evolution affects very significantly the sSFR, size and morphological mix of red 
galaxies, as well as the sSFR of blue galaxies but less significantly. 

However, the timescales involved 
in the consecutive classes in the PPSD are different, in particular, the time from IN to RIN
is short, while that from BS to CL is longer.
It should be kept in mind that not all comparisons between classes are equally trustworthy
according to \citetalias{Coenda:2022}.

\subsection{Quenching vs. morphological transformation, which came first?}

In the previous paragraphs, we showed evidence of the impact of the environment
on star formation quenching, and on morphological transformation. In this subsection, we 
address the 
question of which occurs first. We do this by computing, for each predicted class,
the fraction of passive (i.e. quenched) galaxies and the fraction of early type 
galaxies. In Fig. \ref{fig:comp_frac} we show these fractions as a function of stellar
mass. In this figure, to decide whether a galaxy is passive, we use the 
critical line in the plane $\log(M_{\star})-\log(\mathrm{sSFR})$ of \citet{Knobel:2015}. On the other hand, 
to decide whether a galaxy is early we use the condition $P_{\rm E}>P_{\rm S}$.
We abandon, for the moment, our comfort zone criterion of dealing 
separately with red and blue galaxies. Although contamination and misclassifications 
within each predicted sample are stronger now, they affect the resulting fractions in the same way.

For all predicted classes, in Fig. \ref{fig:comp_frac}, the fraction of passive
galaxies is greater than that of ellipticals over the whole mass range probed for CL, and
over most of the mass range for the other classes.
The intersecting-point between the two fractions shifts towards lower values moving 
from ITL to BS, ranging from $\log(M_{\star}/h^{-1} M_{\odot})\sim 10.3$ 
for ITL, to $\log(M_{\star}/h^{-1} M_{\odot})\sim 9.7$ for BS.
As expected, in general, the fraction of both, quenched and ellipticals galaxies
increases from ITL to CL. A final, more subtle trend in this figure is that the fractions
of passive and ellipticals appear to have different shapes as a function of mass. While the 
fraction
of passive galaxies tends to be a concave function (at least for CL, BS and RIN), the 
fraction of ellipticals is straighter, or even becomes convex at the high mass end.

In addition to having more quenched galaxies than ellipticals 
for the better part of the mass range and for all predicted classes, we observe
that the concave shape of the fraction of quenched galaxies (chiefly for RIN, BS, and CL) shows a tendency of 
being roughly insensitive to mass at $\log(M_{\star}/h^{-1} M_{\odot})\gtrsim 10.5$,
and a sharp decline with decreasing mass at the lower mass end. In contrast, the fraction of ellipticals 
is, in all cases, a function that grows monotonically with mass.

In the lower panels of columns 2--5 of Fig. \ref{fig:comp_frac} we show the
difference of the fractions in the upper panels relative to those for the ITL class.
The differences between IN and ITL are almost negligible over the whole mass 
range. This is not the case for the other classes where some noteworthy features
are seen. Firstly, apart from the highest mass bin, the fraction of passive galaxies is 
clearly higher in all three cases. In increasing order we have BS, RIN, and CL. The swap between
RIN and BS is expected since the predicted class RIN is contaminated by CL galaxies.
Secondly, and most interestingly, the fraction of early type galaxies is consistently different
over the whole mass range only for CL.

These results imply that, typically, it takes a cluster a shorter timescale to quench a 
galaxy than to transform its morphology from spiral to elliptical. 
While an important fraction of galaxies end up quenched after 
diving into a cluster (RIN), or after a single excursion in and out a cluster (BS), it takes 
the full action of the cluster environment over a longer timescale to morphologically change a 
significant fraction of galaxies.

\section{Discussion and conclusions}\label{sect:conclu}

In this paper, we apply the code \roger\ of \citet{delosrios:2021}
on SDSS galaxies located in volumes defined around a sample of bright X-ray
galaxy clusters, to compute each galaxy's probabilities of being of five dynamical
classes: cluster galaxies, recent infallers, backsplash galaxies, infallers, and 
interlopers. \roger\ uses as input the position of each galaxy in the projected 
phase space diagram.
Using these probabilities and the thresholds selected by 
\citet{Coenda:2022}, we classify galaxies in our sample into 
the five predicted categories.
We then compare a number of galaxy properties across the five predicted classes
in order to shed light onto the environmental effects produced by clusters upon
galaxies. 
\roger\ proves to be an important tool to perform this kind of classification. Furthermore, the results of \citet{delosrios:2021} and \citet{Coenda:2022} 
help understand how the resulting samples are contaminated by misclassifications
and how the results of comparing across the five predicted classes have to be interpreted.

We study separately red and blue galaxies, following \citet{Coenda:2022}. 
CL, BS and RIN galaxies are the predicted classes upon which the cluster environment
has left its imprint in different degrees. On the other hand,
IN and ITL galaxies are useful as control samples of galaxies
that have not experienced the environmental effects of clusters. 

For red galaxies, we find that the CL class stand out as different from the other predicted
classes, having a suppressed specific star formation rate, smaller sizes, and
are more likely to be classified as ellipticals. Among the other classes differences
are more subtle. However, we note a significant difference in sSFR between
BS and ITL, and a non negligible difference between BS and IN. This is an
indication that the cluster environment diminishes a galaxy's star formation
after a single passage in agreement with the results of \citet{Mahajan:2011}. 
But it takes more than an excursion through the
cluster to significantly shrink a red galaxy. On the other hand, for blue galaxies, and particularly with $\log(M_{\star}/h^{-1}M_{\odot})
\lesssim 10.5$, we find evidence that RIN are smaller 
than IN and ITL, while the latter two are comparable. 
This may indicate that these blue galaxies are effectively reduced in size 
in their first dive into the cluster. We also find evidence of an increase in the 
probability of these RIN galaxies of being elliptical, and a reduction in their star  
formation rate. 
It should be recalled that SDSS spectroscopy involves the inner regions of
galaxies, i.e., the optical fiber size do not allow for a larger spatial coverage,
thus, differences in sSFR reported above can be even greater. This will be addressed
in a forthcoming paper (Muriel et al. in preparation) using
OmegaWINGS data \citep{Gullieuszik15,Moretti17}.

We find that only virialised galaxies in clusters show a significant morphological evolution. 
Nevertheless, we find that blue galaxies that have just got into clusters may have
already undergone morphological transformations.
In a recent work, \citet{Lopez-Gutierrez:2022} associate spiral galaxies observed in 
the vicinity of clusters with BS. They argue that the timescales needed to undergo 
a morphological transformation exceed by far the typical timescales BS spend within a cluster.
We find a non negligible fraction of ellipticals among our BS, so to be consistent with the 
timescales mentioned by \citet{Lopez-Gutierrez:2022}, they should have been pre-processed in 
other environments before diving into the clusters.
Further analysis on BS galaxies is currently being carried out by Ruiz et al. (in preparation). 

We address the question of whether the quenching timescales are typically shorter
than the morphological change timescales. We find quenching occurs faster for all predicted
dynamical classes. Furthermore, while quenching is evidently enhanced as soon as galaxies 
get into clusters, morphological transformations require galaxies to experience the
action of the physical mechanisms of the cluster for longer periods of time. 
In agreement with our results,  \citet{kelkar:2019} employ observational data from the ESO Distant
Cluster Survey \citep{EDisCS:2005} to argue that star formation in galaxies is suppressed earlier than 
their morphological transformation which happens on a longer time-scale.
On the other hand, our results disagree with those by \citet{Martig:2009}, \citet{Bait:2017} 
and \citet{Sampaio:2022}.
In particular, \citet{Martig:2009}, using hydrodynamical simulations, present the concept of 
morphological quenching, in which star formation is suppressed in synchronization with morphological 
transformation. 
Similar results are found by \citet{Bait:2017} using multi wavelength data in 
galaxies selected in the SDSS catalog. \citet{Sampaio:2022} argue that 
morphological transformation occurs before star formation quenching. They estimate a timescale of $1\, {\rm Gyr}$ 
for the transition of late to early-type morphology, and a timescale of $3\, {\rm Gyr}$ for the star formation 
quenching.
\citet{Mamon:2019}  
argue that the greater extent of spiral galaxies relative to ellipticals suggests that spirals
should transform into S0s and ellipticals within a single orbit. Part of our discrepancies with other
studies may be caused by the lack of S0 morphologies in Galaxy Zoo. It should take longer for spirals to 
transform into ellipticals than to transform into S0s.

\section*{Acknowledgements}
The authors thank the referee, Gary Mamon, for his suggestions and
comments that resulted in a substantial improvement of the paper.
This paper has been partially supported with grants from Consejo Nacional de 
Investigaciones Cient\'ificas y T\'ecnicas (PIPs 11220130100365CO and 11220210100064CO) 
Argentina, the Agencia 
Nacional de Promoci\'on Cient\'ifica y Tecnol\'ogica (PICT 2020-3690), 
Argentina, and Secretar\'ia de Ciencia y Tecnolog\'ia, Universidad Nacional de C\'ordoba, 
Argentina.
MdlR acknowledges financial support from the Comunidad Aut\'onoma de Madrid through the grant SI2/PBG/2020-00005.
\section*{Data availability}
\roger~data underlying this article are available in \textsc{Github} at 
\url{https://github.com/Martindelosrios/ROGER} and in \textsc{zenodo} at  
\url{https://zenodo.org/badge/latestdoi/224241400}. 
Physical properties of clusters used in this work are available at
\url{https://cdsarc.cds.unistra.fr/viz-bin/cat/J/A+A/504/347} and 
\url{https://cdsarc.cds.unistra.fr/viz-bin/cat/J/A+A/564/A85}
Galaxies used in this article were selected from SDSS-DR7.
Specific star formation rates for these galaxies were taken from
\url{http://www.mpa-garching.mpg.de/SDSS/DR7}.



\bibliographystyle{mnras}





\appendix
\section{Assessing the significance of the observed differences between two classes of galaxies}
\label{test_lares}
To estimate the significance of the observed differences between two 
predicted classes of galaxies, $A$ and $B$, in the trend of a given property, $y$, as a 
function of stellar mass, we proceed as \citet{Muriel:2014}. 
They propose a test to assess the rejection probability of the null hypothesis in which 
the distribution of points in the plane $\log(M_{\star})-y$ of the two sets are drawn
from the same underlying distribution. The test is based on the computation
of the normalised cumulative bin-to-bin difference:
\begin{equation}
    \Delta y_{\rm AB} =
    \frac{\sum_{i=1}^{N_\text{bin}}\sum_{j_A=1}^{N_A(i)}\sum_{j_B=1}^{N_B(i)}
    \left[ y_A(j_A)-y_B(j_B)\right]}{\sum_{k=1}^{N_\text{bin}}N_A(k)N_B(k)},
\end{equation}
where $N_\text{bin}$ bins in $\log(M_{\star})$ are used,  
in the $i-$th bin there are $N_A(i)$ galaxies of the class $A$ and $N_B(i)$ galaxies of the class $B$,
each having a value of the $y$ property denoted by $y_A$ and $y_B$, respectively.
The sample of galaxies of class $A$ has a total number of objects $N_A^\text{tot}$, and the
samples of class $B$ galaxies $N_B^\text{tot}$.
This quantity adds up the differences $y_A-y_B$ between all pairs of objects of the classes $A$ and $B$ 
within each bin, then accumulates across bins, and is finally normalised by the total number
of pairs used. Then, the quantity $\Delta y_{\rm AB}$ is computed repeatedly many times over 
random selections of samples of galaxies in the following way: i) mix the 
samples of galaxies $A$ and $B$ together in a parent sample with $N_{\rm AB}=N_A^\text{tot}+N_B^\text{tot}$
objects in total; ii) from this parent sample select at random two subsamples with 
$N_A^\text{tot}$ and $N_B^\text{tot}$ objects; iii) compute $\Delta y_{\rm AB}$ using these two 
subsamples obtaining a value $\Delta y_\text{ran}$; iv) repeat ii)-iii) a large number of times
to obtain a distribution of $\Delta y_\text{ran}$ which should be Gaussian and centered at 0;
v) the rejection probability of the null hypothesis is computed as 
$\mathcal{R}_\text{p}=1-F$, where
$F$ is the fraction of $\Delta y_\text{ran}$ measurements smaller (larger) than 
$\Delta y_{\rm AB}$ if the latter is positive (negative).
We consider two predicted classes $A$ and $B$ are different in the plane
$\log(M_{\star})-y$ if $\mathcal{R}_\text{p}>0.997$, i.e., if the value of 
$\Delta y_{\rm AB}$ is more than $3\sigma$ away from the mean of the
distribution of $\Delta y_{\text{ran}}$.

\bsp	
\label{lastpage}
\end{document}